\documentclass[floatfix,twocolumn,epsf,prb,amsmath,amssymb,showpacs]{revtex4}
\usepackage{graphicx}

\begin{document}

\newcommand{\ud}{\mathrm{d}}
\newcommand{\kvec}[2]{\begin{pmatrix} #1 \\ #2 \end{pmatrix} }
\newcommand{\kvecje}[2]{\bigl( \begin{smallmatrix} #1 \\ #2 \end{smallmatrix} \bigr)}
\newcommand{\rvec}[2]{\begin{pmatrix}#1 & #2 \end{pmatrix} }
\newcommand{\rvecje}[2]{(\begin{smallmatrix} #1 & #2 \end{smallmatrix})}
\newcommand{\matt}[4]{\begin{pmatrix} #1 & #2 \\ #3 & #4 \end{pmatrix} }
\newcommand{\mattje}[4]{\bigl( \begin{smallmatrix} #1 & #2 \\ #3 & #4 \end{smallmatrix} \bigr)}
\newcommand{\vgll}[2]{\left\{ \begin{array}{c} #1 \\ #2 \end{array} \right. }
\newcommand{\vglL}[2]{\begin{equation} \left\{ \begin{align} #1 \\ #2 \end{align} \end{equation} \right. }
\newcommand{\pd}[2]{\frac{\partial #1}{\partial #2}}
\newcommand{\pdd}[2]{\frac{{\partial}^2 #1}{{\partial #2}^2}}
\newcommand{\pdje}[1]{\partial_{#1}}

\title{Dirac and Klein-Gordon  particles in one-dimensional periodic potentials}
\author{Micha\"el Barbier$^{1,*}$, F. M. Peeters$^{1,+}$, P. Vasilopoulos$^{2,**}$,
and J. Milton Pereira$^{1,3,\dagger}$ Jr.
\ \\}
\affiliation{
\ \\
$^{1}$Department of Physics University of Antwerp Groenenborgerlaan 171, B-2020 Antwerpen, Belgium}
\affiliation{
\ \\
$^{2}$Department of Physics, Concordia University, 7141 Sherbrooke Ouest Montreal Quebec, Canada H4B 1R6\\ }
\ \\
\affiliation{$^{3}$Departamento de F\'{\i}sica, Universidade
Federal do Cear\'a, Fortaleza, Cear\'a, $60455$-$760$, Brazil}

\begin{abstract}
We evaluate the dispersion relation for massless fermions, described by the Dirac equation, and for zero-spin bosons, described by the Klein-Gordon equation, moving in two dimensions and in the presence of a one-dimensional periodic potential. For massless fermions the dispersion relation shows a zero gap for carriers with zero momentum in the direction parallel to the barriers in agreement with the well-known "Klein paradox". Numerical results for the energy spectrum and the density of states are presented. Those for fermions are appropriate to graphene in which carriers behave relativistically with the "light speed" replaced by the Fermi velocity. In addition, we evaluate the transmission through a finite number of barriers for fermions and zero-spin bosons and relate it with that through a superlattice.
\end{abstract}
 \maketitle

\section{Introduction}

The recent realization of stable single layer and bilayer carbon crystals (graphene) has aroused considerable interest in the study of their electronic properties \cite{novo3,zhang}. These materials have unusual properties that may allow the development of carbon-based nanoelectronic devices. The behavior of charge carriers in wide single-layer graphene sheets is that of chiral, "relativistic" massless particles  with a "light speed" equal to the Fermi velocity of the crystal ($v_F \approx c/300$) and a gapless linear dispersion close to the $K$ and $K'$ points \cite{zheng,novo3,shara}. One consequence of that is that single-layer graphene displays an unusual quantum Hall effect, in which the quantum Hall plateaus occur \cite{shara} at half-integer multiples of $4\,e^2/h$.  The absence of a gap and the chiral nature of the electronic states, in both single-layer and bilayer graphene, is at the root of phenomena such as the Klein tunneling \cite{klein,been,kat,milton1} which is the perfect transmission of carriers, upon normal incidence,  through a potential barrier. The study of this effect is relevant to the development of future graphene-based devices \cite{milton2}.

From the standpoint of basic research, due to its lower "light speed" graphene can also fulfill the role of a testbed for the investigation of relativistic quantum effects, such as  {\it Zitterbewegung} \cite{zit} and pair creation \cite{crepa}. Moreover, the role of chirality in the electronic properties of massless fermions can  be assessed or further appreciated by contrasting the behavior of chiral, massless fermions with that of non-chiral, massless zero-spin  bosons.

In this work we  study the dispersion relation for two-dimensional fermions, described by the Dirac equation in the presence of a one-dimensional (1D) periodic potential. This can be realized in a periodically gated graphene layer. In doing so we extend the treatment of Ref.\ \onlinecite{mac} which considered motion only along one direction. For a clearer understanding we contrast these results with those for bosons of zero spin that are described by the Klein-Gordon equation \cite{grei}. Here too we include the motion parallel to the barriers, which, to our knowledge, has  not yet been reported. In addition, we briefly present results, in Sec.\ II, for the transmission through a single barrier. In Sec.\ III  we present the dispersion relation and transmission for either superlattice case  and relate the transmission  to that through a single barrier. A summary and concluding remarks follow in Sec.\ VI.

\section{Tunneling through a single barrier}

\subsection{Bosons}
\begin{figure}[ht]
	\begin{center}
		\includegraphics[width=6cm]{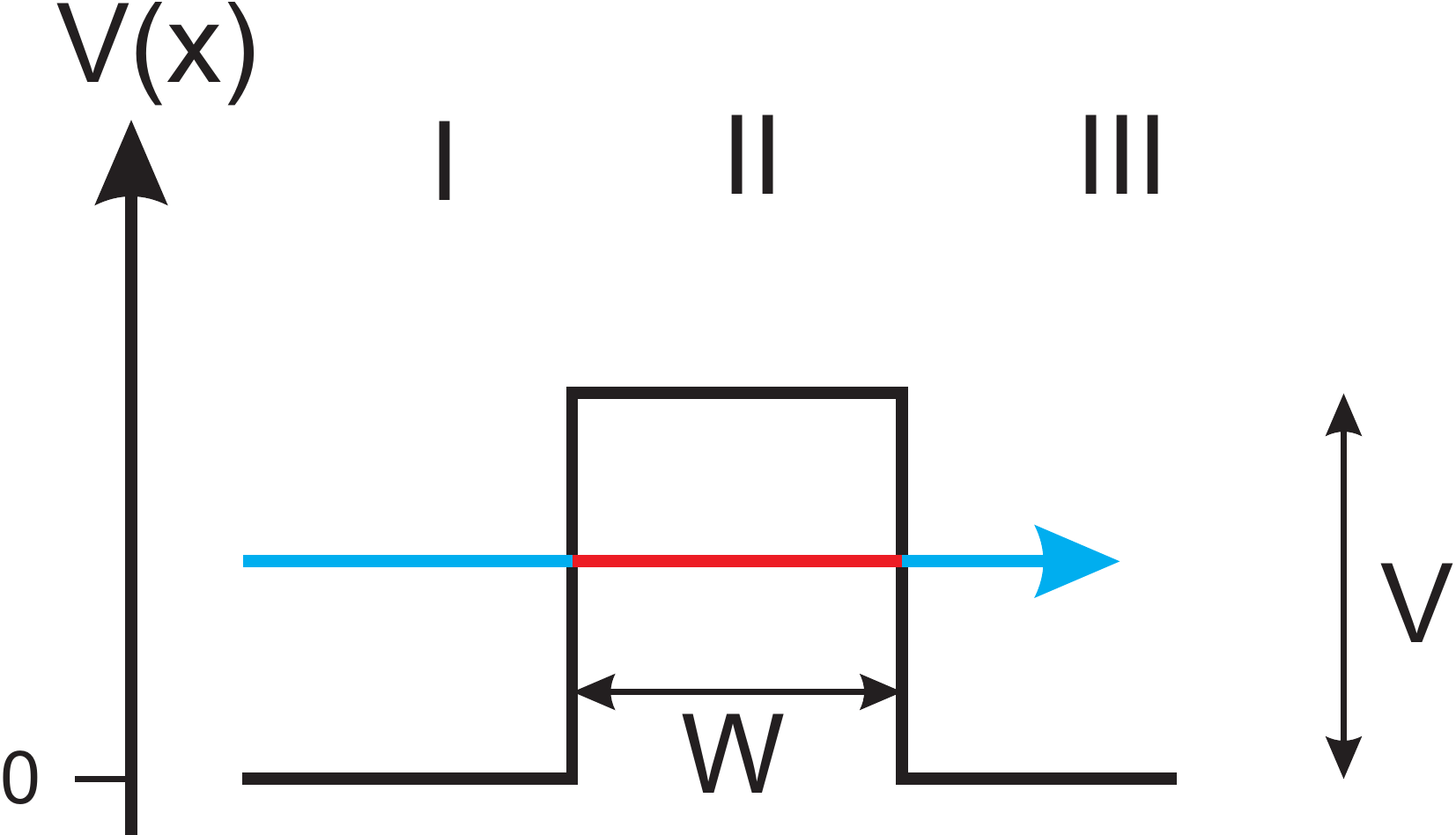}
	\end{center}
	\vspace*{-0.5cm}
	\caption{(Color online) 1D Potential barrier V(x) of height V and width W}\label{fig1}
\end{figure}
We first calculate the transmission of zero-spin bosons through a single barrier of height $V$ and width $W$, shown in Fig.\ \ref{fig1}, using the Klein-Gordon equation in two dimensions
\begin{equation} \label{eq_KG-eq}
	\nabla^2 \psi(x,y) = - \frac{1}{\hbar^2c^2} \left[(E - V(x))^2  - m^2c^4 \right] \psi(x,y).
\end{equation}
Since the hamiltonian $H$ commutes with $p_y$ we look for solutions in the form $\psi(x,y) = \psi(x) e^{ik_yy}$. With this substitution the resulting equation for $\psi(x)$ is solved by $\psi(x) = e^{\pm ik_xx}$ outside the barrier  with $k_x = E'/\hbar c$ and by $\psi(x) = e^{\pm iKx}$ inside it with
\begin{equation}
	K = \frac{1}{\hbar c} [ E'^2 - 2V(E'^2 + \hbar^2 k_{y}^2 c^2 + m^2c^4)^{1/2} + V^2]^{1/2},
\end{equation}
and with $E'$ given by
\begin{equation}
    E'^2 = E^2 - \hbar^2 k_{y}^2 c^2 - m^2c^4 = \hbar^2 k_{x}^2 c^2.
\end{equation}
Then following the standard procedure of matching the solutions and their derivatives at the interfaces of the regions I and II, and II and III, shown in Fig.\ \ref{fig1}, leads to the transmission
\begin{equation}
	\mathcal{T}(k_{x},k_{y}) =
	\left[1 + \left(\frac{k_x^2 - K^2}{2k_xK}\right)^2 \sin^2(K W)\right]^{-1}.
\end{equation}
If $K$ becomes imaginary, then $K$ is replaced by $ i|K|$ and $\sin(KW)$ by $i\sinh(KW)$ in this equation. Notice that in contrast to the non-relativistic case, $K$ depends on $k_y$. The result for the non-relativistic case is obtained by inserting $k_y=0$, $k_x = [2mE/\hbar^2]^{1/2}$, and $K = [2m(E-V)/\hbar^2]^{1/2}$ in Eq.\ (4).

\begin{figure}[ht]
	\begin{center}
	\hspace*{-0.5cm}
    	\includegraphics[width=9cm]{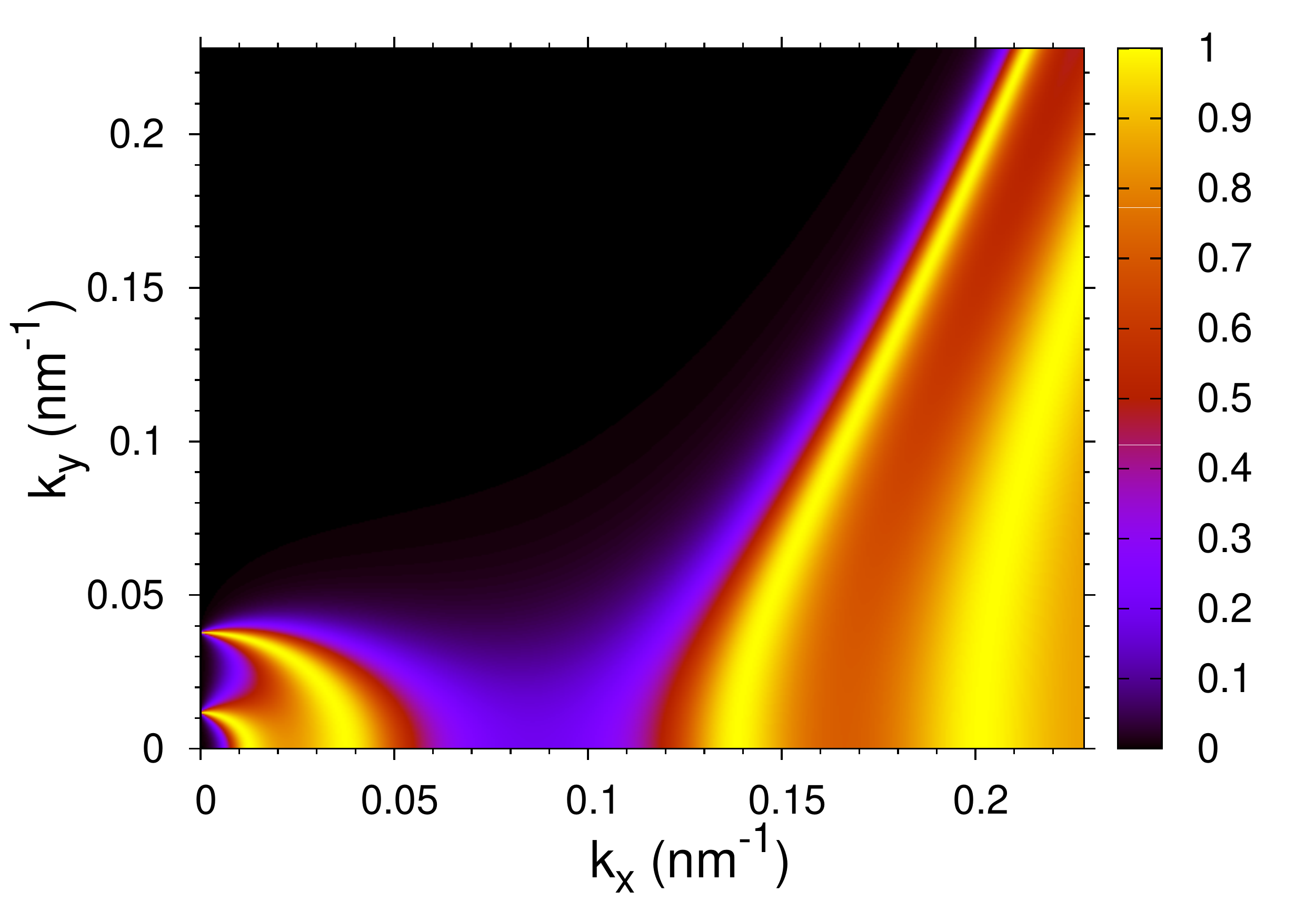}
	\end{center}
	\vspace*{-0.2cm}
	\caption{(Color online) Contour plot of the transmission $\mathcal{T}$ of relativistic bosons through a barrier as a function of $k_x$ and $k_y$ for  $V=50$ meV, $c = 10^6$ m/s, $W = 50$ nm, and  $m=0$.}\label{fig2}
\end{figure}

In Fig.\ \ref{fig2} the transmission through a single potential barrier is plotted for zero-spin bosons with the same velocity ($c = 10^6$m/s) and parameters for the potential barrier as for the massless fermions in graphene. In doing so we will be able to better compare the transmission for bosons with that for fermions. As shown in Fig.\ \ref{fig4}, in which we plot slices of this plot at constant $k_y$, the transmission is perfect for certain values of $k_x$ although the boson has an energy that is much lower than the height of the potential barrier V (if $k_y = 0$ then $E = V = 50$meV for $k_x = 0.076$nm$^{-1}$). Notice that these transmission resonances below the barrier height do not occur in the non-relativistic case. Moreover, the transmission depends on the wavevector component parallel to the barrier, i.e., on $k_y$ since $K$ does, cf.\ Eq.\ (2), whereas it does not in the non-relativistic case.

Since $K$ is always real for $m=0$, the sine term in Eq.\ (4) only introduces a modulation of the transmission. The minima in the transmission arise from the $(k_x^2-K^2)/2k_xK$ term. This is because at the vicinity of $E = V$, the solutions inside the barrier approach those  of a free particle with zero energy and thus with $k=0$, i.e. infinite wavelength. Together with the continuity of $\Psi$ and of its derivative,  this  acts to reduce the amplitude of the transmitted wave. For the non-relativistic case and $E < V$, $K$ becomes imaginary and the sine-term becomes a hyperbolic sine, which can become quite large and reduce significantly the total transmission.

\subsection{Fermions}

In two dimensions spin-1/2 fermions are described by the 2D Dirac equation:
\begin{equation} \label{eq_D-eq}
	\left[ c(\vec{\sigma} \cdot \hat{p}) + m c^2 \sigma_z \right] \Psi = (E - V) \Psi,
\end{equation}
where $\vec{\sigma} = (\sigma_x, \sigma_y)$ and $\sigma_z$ are the Pauli-matrices. Instead of imposing a certain form of solution, without giving any details, to Eq.\ (5) \cite{kat}, we look for solutions in the form $\psi(x,y) = \psi(x) e^{ik_yy}$, where the wave function $\psi(x)$ is a two-component spinor $\psi(x) =[\psi_u(x), \psi_l(x)]^T$. Substituting this form of solution and setting  $E' = E - V(x)$ gives
\begin{equation}
\begin{aligned}
	\partial\psi_l(x)/\partial x + k_y \psi_l(x) & = \frac{i}{\hbar c}(E'-mc^2)\psi_u(x),\\
	\partial \psi_u(x)/\partial x - k_y \psi_u(x) & = \frac{i}{\hbar c}(E'+mc^2)\psi_l(x).
\end{aligned}
\end{equation}
These coupled equations admit solutions of the form
\begin{equation} \label{eq_def_psi_x}
	\psi(x) = \kvec{u_u}{u_l}e^{ik_x x},
\end{equation}
where $k_x = (1/\hbar c)[E^2 - \hbar^2 k_y^2 c^2 - m^2 c^4]^{1/2}$ must be replaced by  $K =(1/\hbar c)[(E-V)^2 - \hbar^2 k_y^2 c^2 - m^2 c^4]^{1/2}$  inside the barrier. Then substituting this form in Eq.\ (\ref{eq_D-eq}) gives
\begin{equation}
    \matt{E' - mc^2}{- \hbar c k_{x,-}}{- \hbar c k_{x,+}}{E' + mc^2} \kvec{u_u}{u_l} = 0,
\end{equation}
where $k_{x,\pm}=k_x \pm ik_y$. A non-trivial solution demands a zero determinant and this leads to
\begin{equation}
    E'^2 = \hbar^2c^2k_x^2 + \hbar^2c^2k_y^2 + m^2c^4
\end{equation}
and to
\begin{equation}
\begin{aligned}
    u_l & = \frac{\hbar c k_{x,+}} {E' + mc^2} u_u, \qquad  E' > 0\\
    u_u & = \frac{-\hbar c k_{x,-}} {|E'| + mc^2} u_l, \qquad E' < 0.
\end{aligned}
\end{equation}
As detailed in the Appendix, the general solution is a linear combination of two independent solutions $\phi_a(x)$ and $\phi_b(x)$ of this form. In contrast with the  boson case of Sec.\ II. A, the continuity of the derivative of the spinor components for Dirac fermions is not necessary. Thus, matching only the solutions at the interfaces between the regions I and II and II and III gives, for $E>V$, the transmission as
\begin{equation} \label{eq_trans_D-eq}
    \mathcal{T}(k_x, k_y) = \left[1+\left( \frac{F}{4k_x^2K^2}-1\right) \sin^2(KW)\right]^{-1},
\end{equation}
where
\begin{equation}
	F=\left(k_x^2\frac{D}{d} + K^2\frac{d}{D} +	 k_y^2(\frac{d}{D}+\frac{D}{d}-2)\right)^2,
\end{equation}
with $d = (|E| + m c^2)/\hbar c$ and $D = (|E-V| + m c^2)/\hbar c$. Similar expressions are obtained for $0 < E < V$ and $E < 0$, cf.\ Appendix.

\begin{figure}[ht]
	\begin{center}
	\hspace*{-0.5cm}
    \includegraphics[width=9cm]{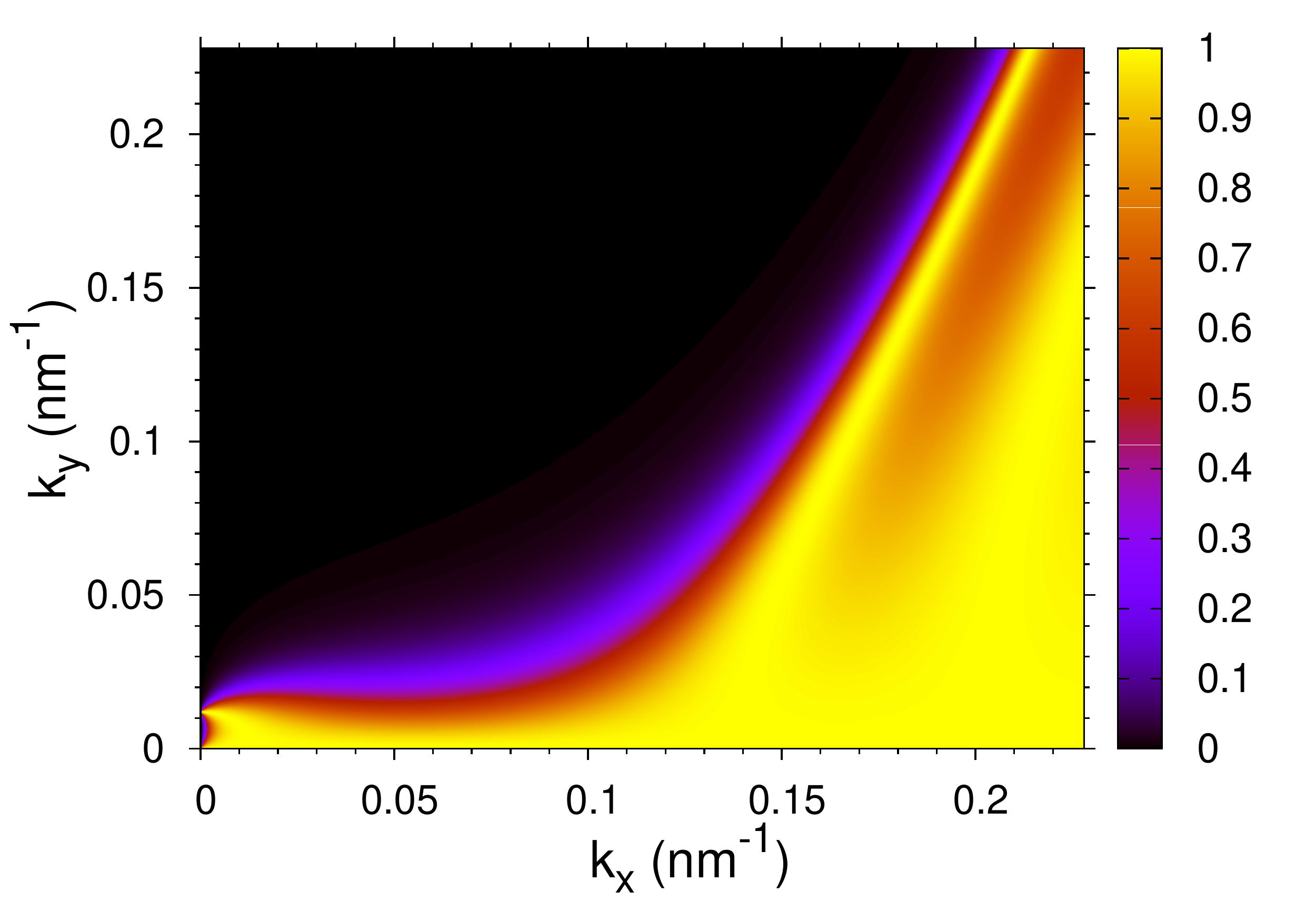}
    \end{center}
	\vspace*{-0.3cm}
	\caption{(Color online) Contour plot of transmission $\mathcal{T}$ of massless relativistic electrons through a barrier with $m=0$, $V=50$ meV, $c = 10^6$ m/s, $W = 50$nm}\label{fig3}
\end{figure}

\begin{figure}[ht]
	\begin{center}
    \includegraphics[width=9cm]{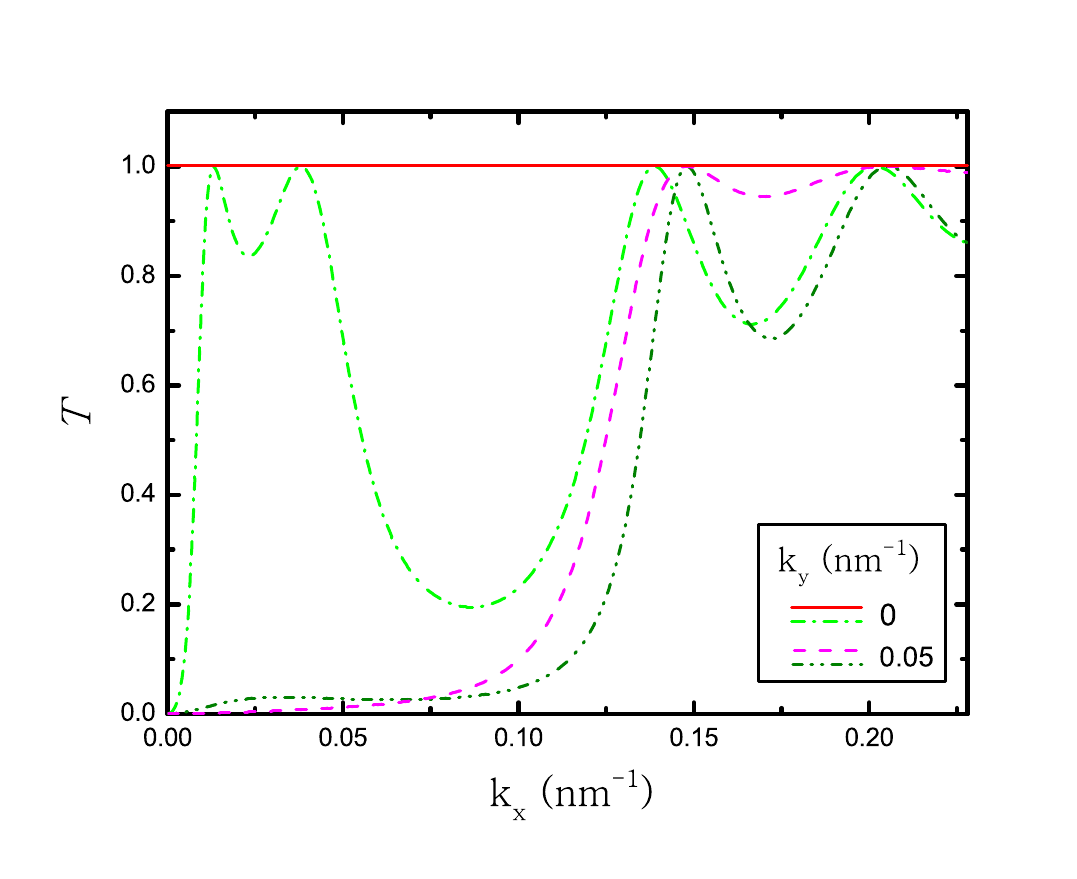}
    \end{center}
	\vspace*{-0.4cm}
	\caption{(Color online) Plot of slices of the transmission coefficient for bosons (green curves) and electrons (solid red and dashed magenta curves) taken at constant $k_y = 0$ and $k_y=0.05$ nm$^{-1}$ from Figs.\ \ref{fig2} and \ref{fig3}, respectively.}
	\label{fig4}
\end{figure}

For $|E - V|^2 < \hbar^2k_y^2c^2 + m^2c^4$, $K$ is imaginary and the transmission is given by Eq.\ (\ref{eq_trans_D-eq}) with $K$ replaced by $iK'\equiv i|K|$ and $\sin(KW)$ by $i\sinh(KW)$. A contour plot of  the transmission is given in Fig.\ \ref{fig3}. As can be seen, for normal incidence, i.e., for $k_y=0$, the transmission is equal to $1$. Contrasting Fig.\ \ref{fig3} with Fig.\ \ref{fig2} for bosons we see a significant gap in Fig.\ \ref{fig2} for normal incidence that is not present in Fig.\ \ref{fig3}. Another way to contrast the results of the two cases is shown in Fig.\ \ref{fig4} where slices of the transmission across the planes $k_y = 0$ and $k_y =0.05$ nm$^{-1}$ are shown for bosons (green  curves) and electrons (red and magenta curves). The perfect transmission ($\mathcal{T}(k_x, k_y)=1$) for normal incidence results from the chirality of the carriers which prevents them from being reflected by a barrier, see also Ref.\ \onlinecite{kat}. Because massless bosons are not chiral, they are not subject to such a constraint.

\section{Superlattices}
We now consider 1D superlattices (SL), that are periodic structures consisting of $N$ identical units of length $L$. Each unit consists of a rectangular barrier and a rectangular well whose widths are denoted, respectively, by $a$ and $b$. We first consider fermions described by the Dirac equation, Eq.\ (\ref{eq_D-eq}). For motion of electrons along only one direction (x), such a calculation was done by McKellar {\it et al.} \cite{mac}. We generalize this earlier work to two dimensions.

We start with Eqs. (\ref{eq_D-eq})-(\ref{eq_def_psi_x})  taking $c \approx 10^6$m/ for our numerical results, a value that is appropriate for graphene. Equation (\ref{eq_def_psi_x}) can be written as
\begin{equation} \label{psi1}
    \psi(x) = \kvec{\psi_u(x)}{\psi_l(x)}.
\end{equation}
Using Eq.\ (\ref{eq_D-eq}) we have two independent solutions $\phi_a(x)$ and $\phi_b(x)$ in each unit cell. The solution in the $n$th unit cell $n$,  written as $\psi_n(x) = \psi(x + (n-1)l)$, is a linear combination of these two
\begin{align} \label{fi1}
	\hspace*{-0.1cm}
    \psi_n(x) = a_n \kvec{\phi_{a,u}(x)}{\phi_{a,l}(x)} + b_n \kvec{\phi_{b,u}(x)}
    {\phi_{b,l}(x)}, \,\,  x \in [0,l].
\end{align}
It is convenient to define the matrix
\begin{equation} \label{ab3}
    {\bf\Omega}(x) = \matt{\phi_{a,u}(x)}{\phi_{b,u}(x)}{\phi_{a,l}(x)}{\phi_{b,l}(x)}.
\end{equation}
Then the continuity of the wave function leads to the following relation between the coefficients of the unit cells:
\begin{align} \label{ctu2}
    {\bf\Omega}(0) \kvec{a_{n+1}}{b_{n+1}} & = {\bf \Omega}(l) \kvec{a_{n}}{b_{n}}\\
    \Rightarrow \kvec{a_{n+1}}{b_{n+1}} & = {\bf T} \kvec{a_{n}}{b_{n}},
\end{align}
and  the transfer matrix ${\bf T}$ is given by ${\bf T} = {\bf\Omega}(0)^{-1} {\bf\Omega}(l)$.

It can be proven \cite{mac} that $\det[{\bf T}] = 1$. Assuming periodic boundary conditions we can write
\begin{align} \label{per5}
    \psi(x+L) & = \psi(x),\\
    \Rightarrow \kvec{a_{N+n}}{b_{N+n}} & = \kvec{a_n}{b_n},
\end{align}
and obtain
\begin{equation} \label{per6}
    \kvec{a_{N+n}}{b_{N+n}} = {\bf T}^N \kvec{a_n}{b_n},\\
\end{equation}
\begin{equation} \label{per7}
    \Rightarrow {\bf T}^N = 1.
\end{equation}
It follows that the eigenvalues of ${\bf T}$ are $e^{2 \pi i n/N}$, with $n$ an integer, and  its determinant is equal to $1$. The dispersion relation is then given by the trace of ${\bf T}$ as
\begin{equation} \label{ek5}
    2 \cos(k_x l) = Tr({\bf T}),
\end{equation}
where $k_x = 2 \pi n/L$. Notice that this $k_x$, or Bloch wavevector, expresses the periodicity of the structure and should not be confused with the $k_x$ of Sec.\ II.

\subsection{1D Superlattices in graphene}

The dispersion relation is given by Eq.\ (22) and the transfer matrix
${\bf T}$  by
\begin{equation} \label{S}
    {\bf T} = {\bf \Omega}_\kappa (l) {\bf\Omega}_\kappa (a)^{-1} {\bf\Omega}_K(a)
    {\bf \Omega}_K(0)^{-1},
\end{equation}
where ${\bf \Omega}_K$ and ${\bf \Omega}_\kappa$ are the matrices in the barriers and wells, respectively. With the help of the Appendix  we can specify the matrices ${\bf \Omega}_K$ and ${\bf \Omega}_\kappa$. For $E>V$ the dispersion relation becomes ($\kappa=E'/\hbar c$)
\begin{equation} \label{ek2D1}
    2 \cos(k_x l) = 2 \cos(Ka)\cos(\kappa b)-G \sin(Ka) \sin(\kappa b),
\end{equation}
where
\begin{equation}
	G = \left( \frac{\kappa^2 + k_y^2}{d^2} + \frac{K^2 + k_y^2}{D^2}\right) \frac{dD}{\kappa K} -\frac{2k_y^2}{\kappa K}.
\end{equation}
By setting $k_y=0$ in Eqs. (24) and (25) we recover the result of Ref.\ \onlinecite{mac} for 1D motion of Dirac particles. For $(E - V)^2 < m^2c^4 + \hbar^2k_y^2c^2$ and $E - V > 0$ we have an imaginary $K$ and simply make the substitution $K \rightarrow i|K|$ in Eq.\ (24). For $E < 0$  the result for $G$ is the same whereas for   $V> E > 0$ it becomes
\begin{equation}
	G = - \frac{dD}{K\kappa} - \frac{(K^2+k_{y}^2)(\kappa^2+k_{y}^2)}{K\kappa dD}-\frac{2k_y^2}{\kappa K}.
\end{equation}

\begin{figure}[ht]
	\begin{center}
	\includegraphics[width=9cm]{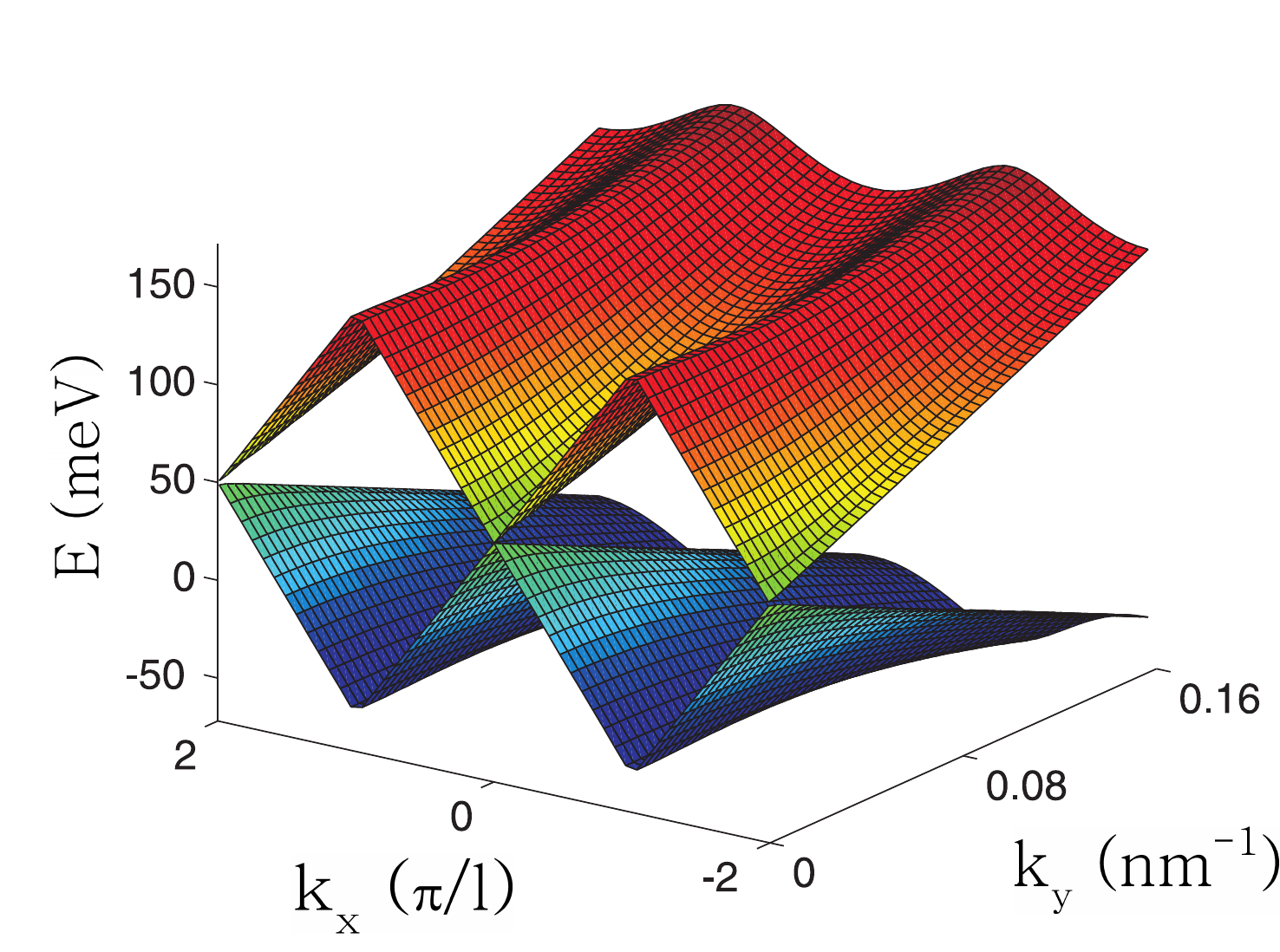}
    \end{center}
	\vspace*{-0.3cm}
	\caption{(Color online) Dispersion relation in a 1D graphene SL. Only the first two bands are shown for $a = b = 10$ nm, $V = 100$ meV, and $m = 0$.}	\label{fig5}
\end{figure}

The first two minibands of the dispersion-relation for a 1D graphene-based SL are plotted in Fig.\ \ref{fig5}. Notice that the form of the dispersion relation in the $y$ direction, in contrast with the non-relativistic case, depends on the $x$ component $k_x$ of the wave vector and is not the energy of a free particle. As a result, we cannot split the energy of the minibands in the manner $E_n = E_n(x) + c\hbar k_y$. If $k_y$ is sufficiently large, this dependence goes away and we have again a linear relation between the energy and $k_y$, independent of the $k_x$ component. Notice also that for $k_y=0$ (and $m=0$) the linear dispersion relation $E(k_x,0)=v_F\hbar k_x$ is unaltered by the periodic potential $V$, it  is only shifted by $V/2$. This is a consequence of the Klein paradox.

\begin{figure}[ht]
	\begin{center}
		\includegraphics[width=7cm]{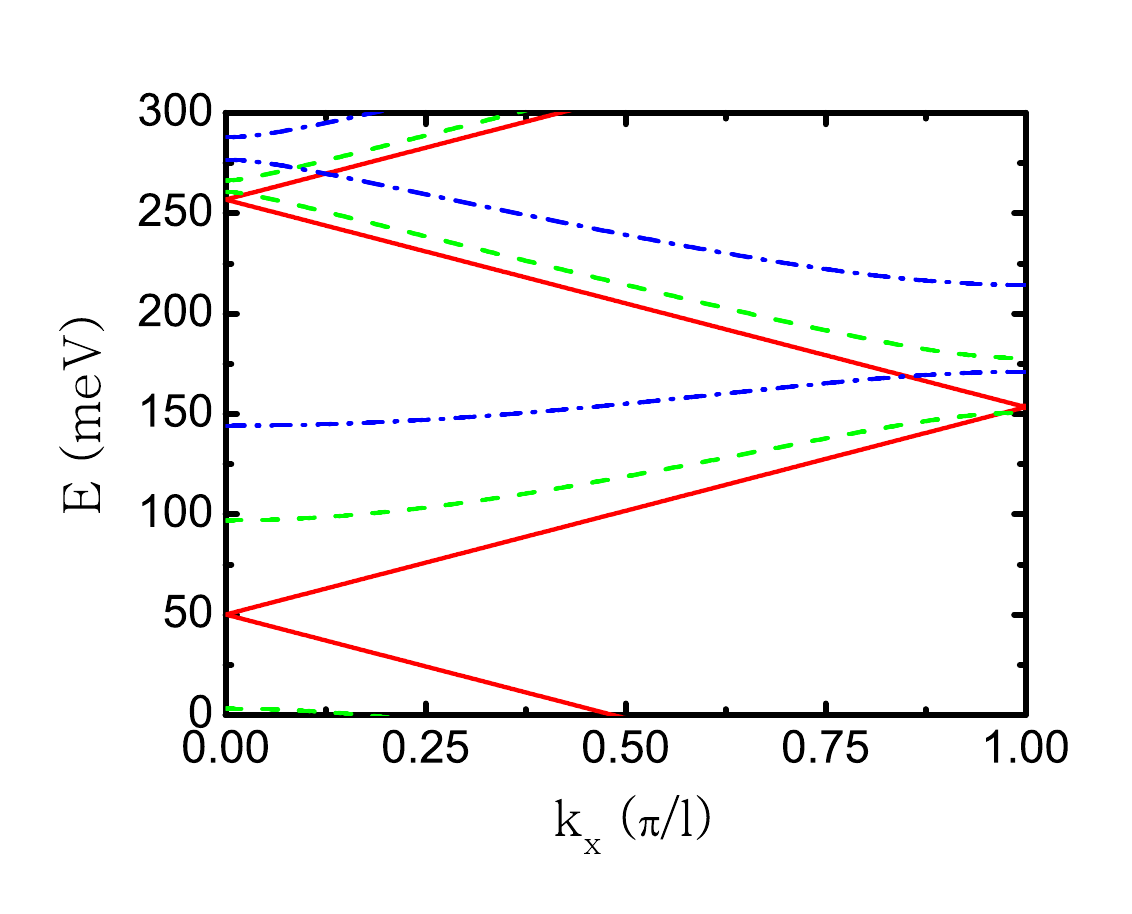}
		\includegraphics[width=7cm]{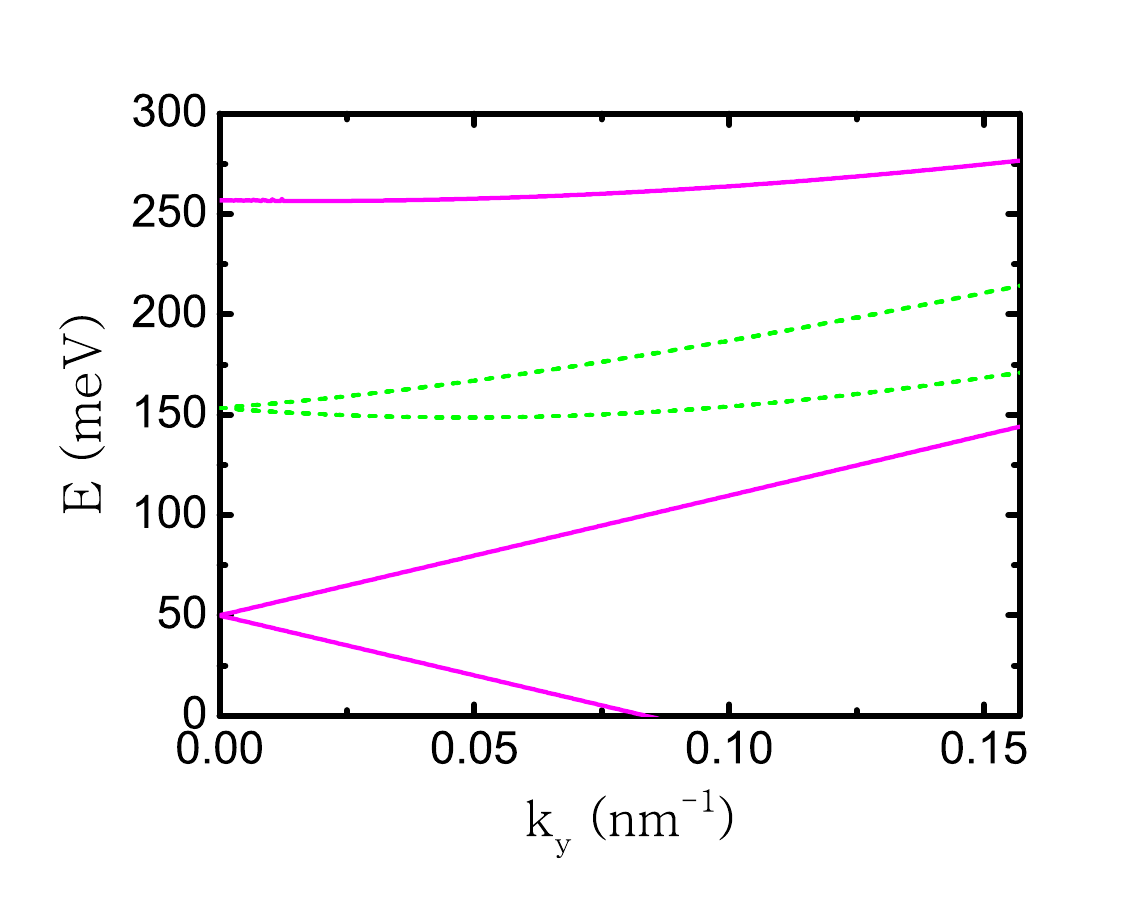}
    \end{center}
	\vspace*{-0.3cm}
    \caption{(Color online) Slices of the dispersion relation shown in Fig.\ \ref{fig5}. (a) is for constant $k_y = 0,\,\, 0.066,\,\, 0.132$ nm$^{-1}$ for the solid red,  dashed green, and dash-dotted  blue  curves, respectively, and (b) for constant $k_x = 0,\,\,\pi/l$ for the solid magenta and dashed green curves, respectively.} \label{fig6}
\end{figure}

In Fig.\ \ref{fig6} we show three slices of the dispersion relation of Fig.\ \ref{fig5}, taken at constant $k_y = 0,\,\, 0.066,\,\, 0.132$ nm$^{-1}$ (red solid, green dashed, and blue dash-dotted curves, respectively). We see that the $k_y$ component is also generating a band gap.

It is instructive to take the Kronig-Penney limit of the dispersion relations given above. For very thin and high barriers, such that $a \rightarrow 0$ and $V \rightarrow \infty$ but with their product ($S = a V$) kept a constant we have
\begin{equation}
\begin{aligned}
	\hspace*{-0.1cm}
  \kappa &= \frac{\sqrt{E^2 - \hbar^2c^2ky^2 - m^2c^4}}{\hbar c},\\
	K &= \frac{\sqrt{(E-V)^2 - \hbar^2c^2ky^2 - m^2c^4}}{\hbar c} \approx \frac{V}{\hbar c},
\end{aligned}
\end{equation}
\begin{equation}
    \frac{\kappa}{d} = \frac{\sqrt{E^2 - \hbar^2c^2ky^2 - m^2c^4}}{|E| + m c^2}, \quad \frac{K}{D} = \frac{K\hbar c}{|E'| + m c^2} \approx 1.
\end{equation}
For $V \rightarrow \infty$ only the cases with positron waves in
the barrier are allowed. Then the dispersion relation becomes

1) $|E| > mc^2$
\begin{equation} \label{ek1D4}
    2\cos(k_x l) = 2 \cos(\frac{S}{\hbar c})\cos(\kappa l) +
    G_+ \sin(\frac{S}{\hbar c}) \sin(\kappa l),
\end{equation}

2) $|E| < mc^2$
\begin{equation} \label{ek1D5}
	\hspace*{-0.1cm}    2\cos(k_x l) = 2 \cos(\frac{S}{\hbar c})\cosh(\kappa l) +
	G_- \sin(\frac{S}{\hbar c}) \sinh(\kappa l),
\end{equation}
where $G_\pm=\kappa/d \pm d/\kappa + k_y^2/\kappa d$.

\begin{figure}[ht]
  \begin{center}
   \includegraphics[width=9cm]{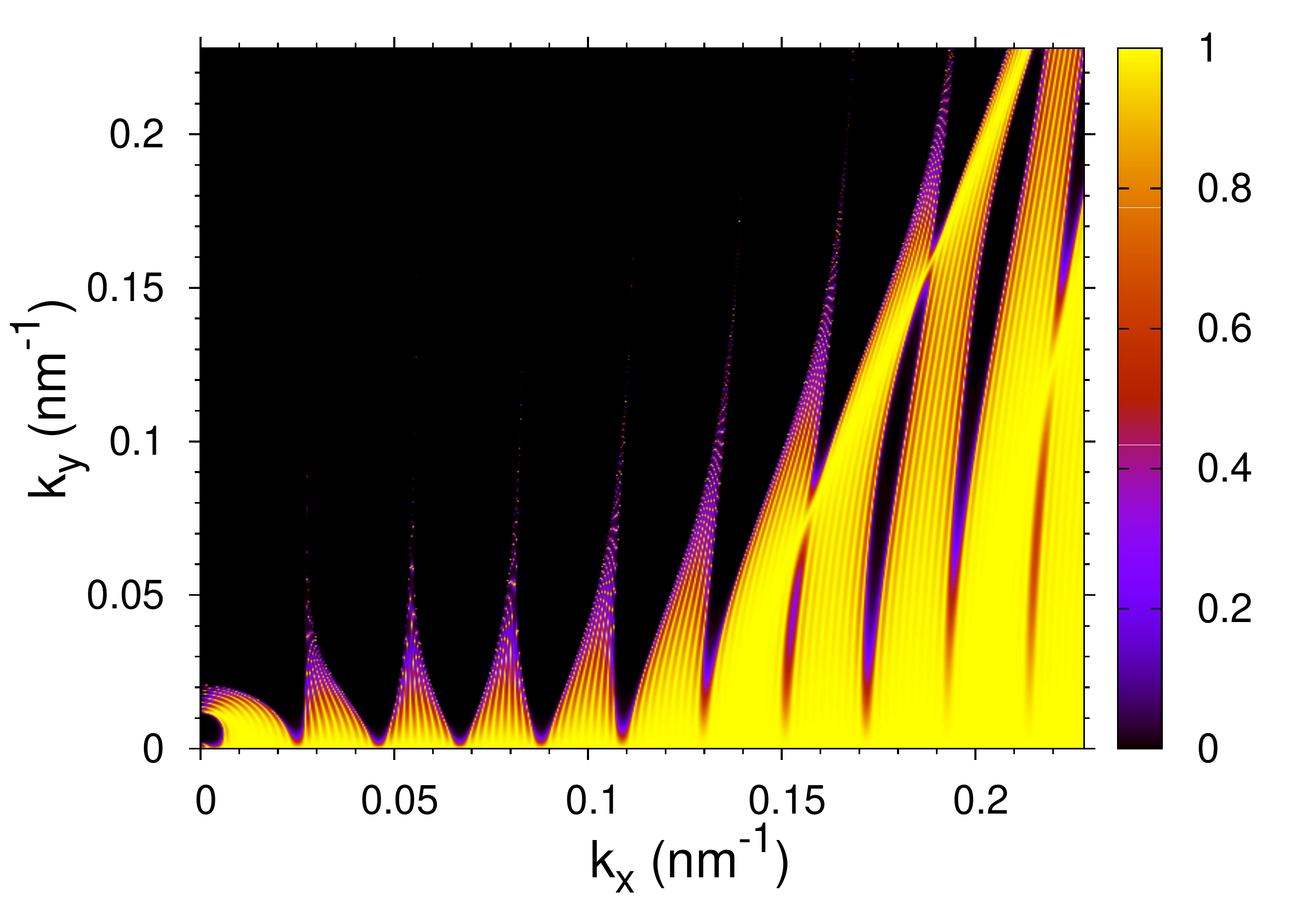}
    \end{center}
\vspace*{-0.4cm}
  \caption{(Color online) Contour plot of the transmission $\mathcal{T}$ of relativistic fermions in a structure consisting of ten units with $c = 10^6$ m/s, $V = 50$ meV, $a =50$ nm, $ b = 100$ nm, and $m = 0$.} \label{fig7}
\end{figure}

To make contact with Sec.\ II, we now consider the transmission through a multi-barrier structure that is a good approximation to a SL. In Fig.\ \ref{fig7} we show a contour plot of the transmission for a  structure consisting of ten units. Upon contrasting Figs.\ \ref{fig3} and \ref{fig7} it is seen that the main effect of having many units is to slightly reduce the continuous range of $k_x$ for which a perfect transmission occurs by "inserting" small gaps between  regions of perfect transmission. The resonances also show a very slight increase in the transmission. This behavior remains essentially unaltered if the number of units is increased and is similar to that reported previously in the context of resonant tunneling through a double barrier \cite{milton2}.

\subsection{Bosons}
The dispersion relation for a SL and relativistic bosons is obtained in a similar way and in full analogy with the non-relativistic Schr\"odinger equation. We use the Klein-Gordon equation, Eq.\ (\ref{eq_KG-eq}), and obtain the same dispersion relation as in the non-relativistic case but with different values for $ \kappa$ and $K$, namely,
\begin{align}
	& \kappa = (1/\hbar c) [E^2 -\hbar^2k_y^2c^2 - m^2c^4]^{1/2}, \\
	& K = (1/\hbar c) [(E - V)^2 - \hbar^2 k_{y}^2 c^2 - m^2c^4]^{1/2}.
\end{align}
The dispersion relation is
\begin{equation}
	2 \cos(k_x l) = 2 \cos(Ka)\cos(\kappa b) - \frac{\kappa^2 + K^2}{\kappa K} \sin(Ka) \sin(\kappa b).
\end{equation}
If  $K$ becomes imaginary, which happens for $(E-V)^2 < m^2c^4 + \hbar^2k_y^2c^2$, we simply replace $K$ by $i|K|$  in Eq.\ (34).

\begin{figure}[ht]
	\begin{center}
	\includegraphics[width=9cm]{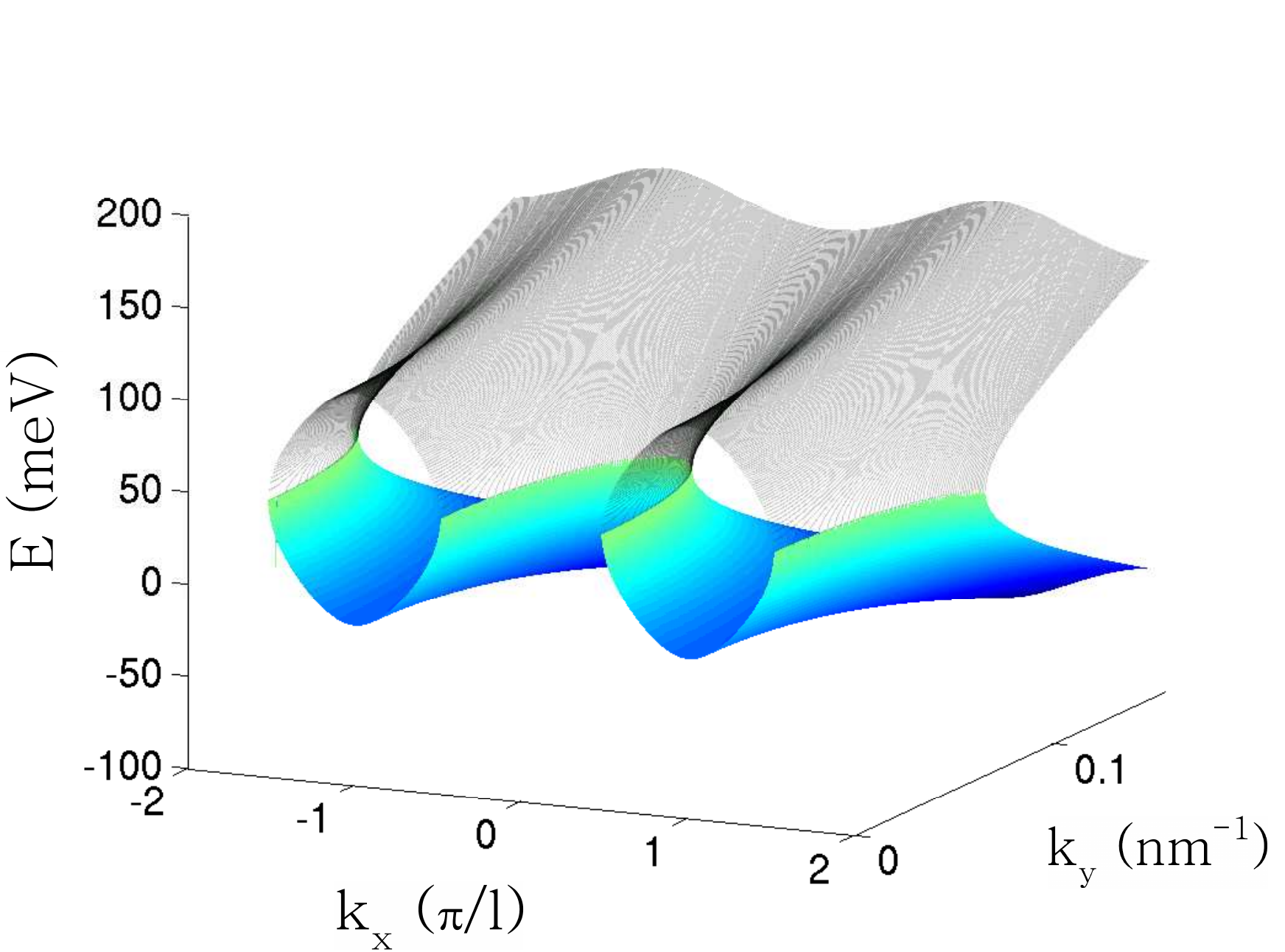}
	\end{center}
	\vspace*{-0.4cm}
	\caption{(Color online) Dispersion relation for relativistic massless zero-spin bosons.  Only the first two minibands are shown  for $c = 10^6$ m/s, $V = 100$ meV, $a = b = 10$ nm, and $m = 0$.} \label{fig8}
\end{figure}

In Fig.\ \ref{fig8} we plot the first two minibands as a function of $k_x$ and $k_y$ for the parameters shown in the caption. The main difference with the non-relativistic case is the appearance of a large gap along $k_x$ and part of $k_y$, centered around $k_x=0$. Notice the turning point of the band, which leads to an infinite group velocity $v_{k_y}$. Such an unusual result is not uncommon in connection with the Klein-Gordon equation\cite{ste}. If we take the mass sufficiently different from zero the turning point in the superlattice band structure disappears. Notice the similarity and differences with the relativistic fermion case depicted in Fig.\ \ref{fig5}. The situation is similar between Figs.\ \ref{fig6} and \ref{fig6} where  the dispersion relation is shown vs $k_x$ in the left panels and vs $k_y$ in the right ones.
\begin{figure}[ht]
\begin{center}
	\includegraphics[width=7cm]{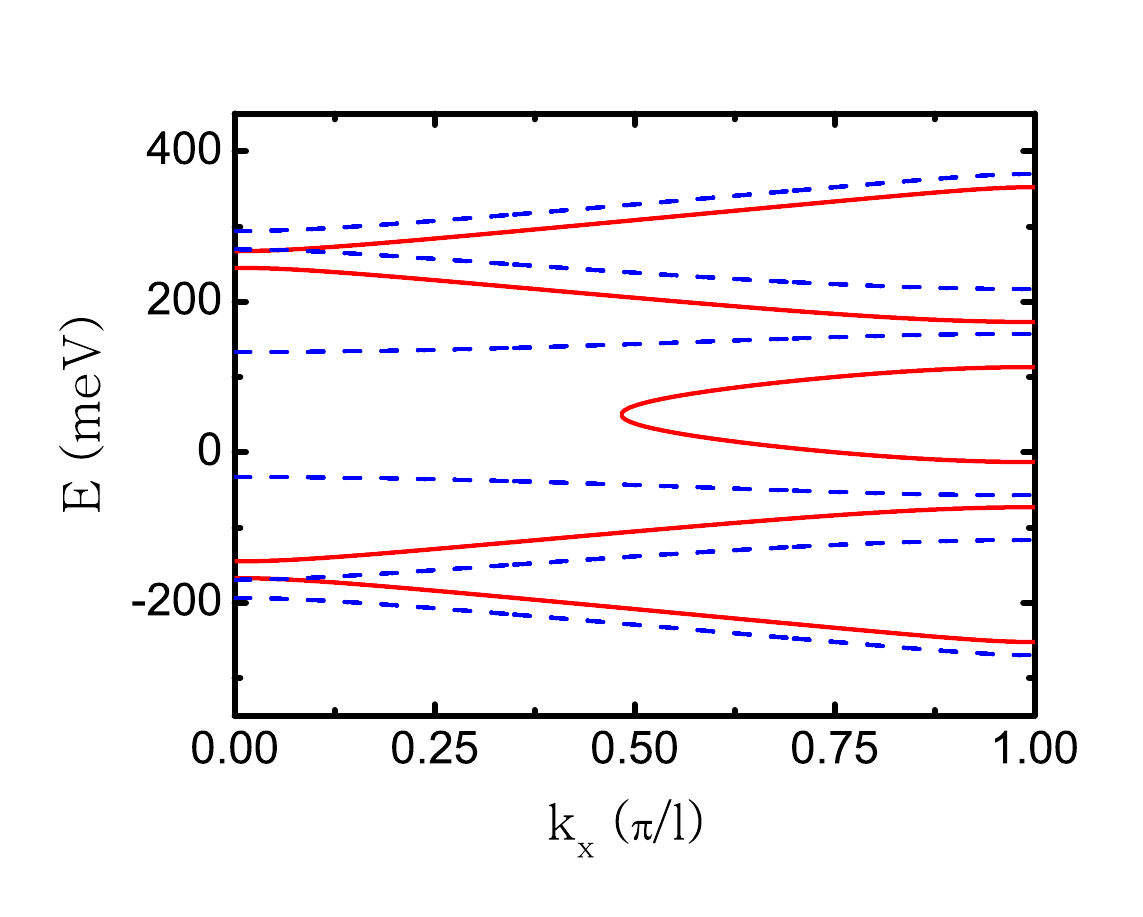}
	\includegraphics[width=7cm]{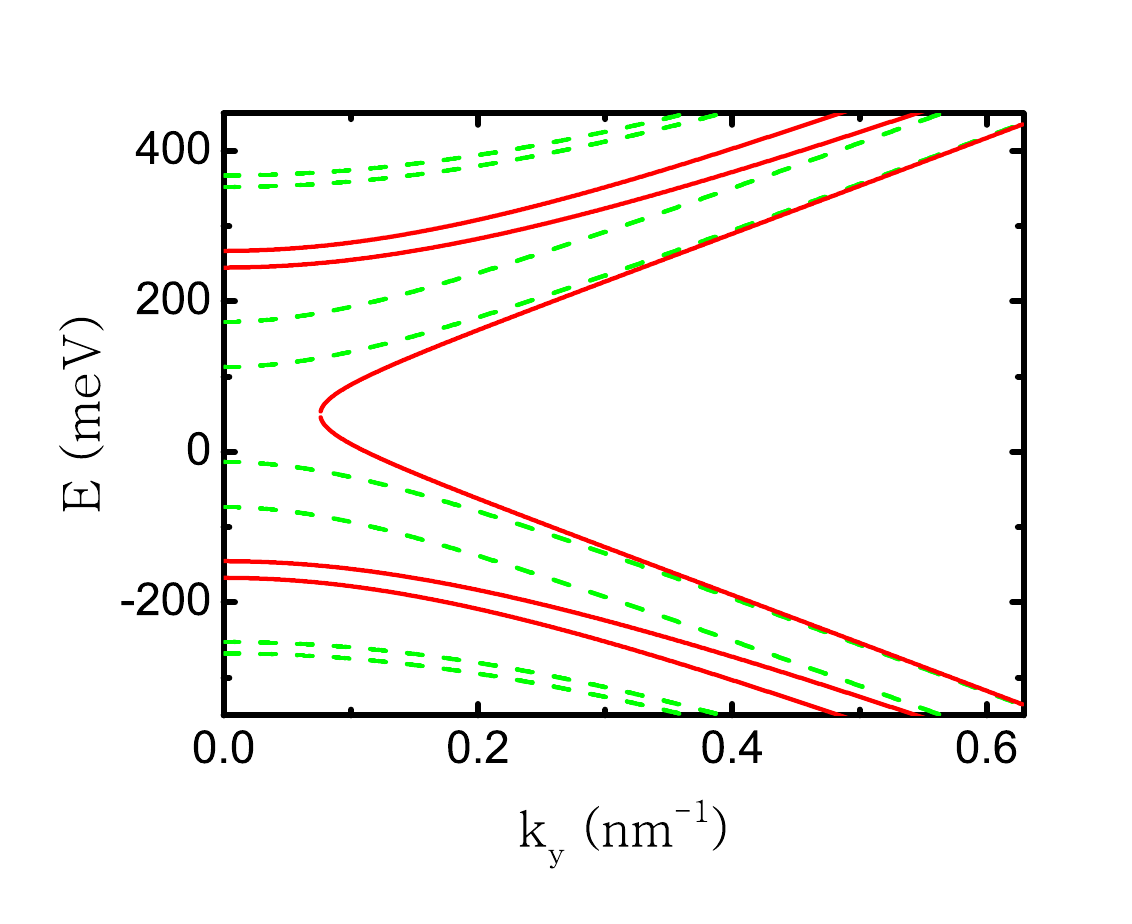}
	\end{center}
	\vspace*{-0.5cm}
	\caption{(Color online) Slices of the dispersion relation for relativistic bosons extracted from Fig.\ \ref{fig8}. (a) is for constant $k_y=0, 0.16$ nm$^{-1}$, for the solid red and dashed blue curves, respectively, and (b) for constant $k_x = 0, \pi/\ell$ for the solid red and dashed green curves, respectively.} \label{fig9}
\end{figure}

\begin{figure}[ht]
	\begin{center}
		\includegraphics[width=9cm]{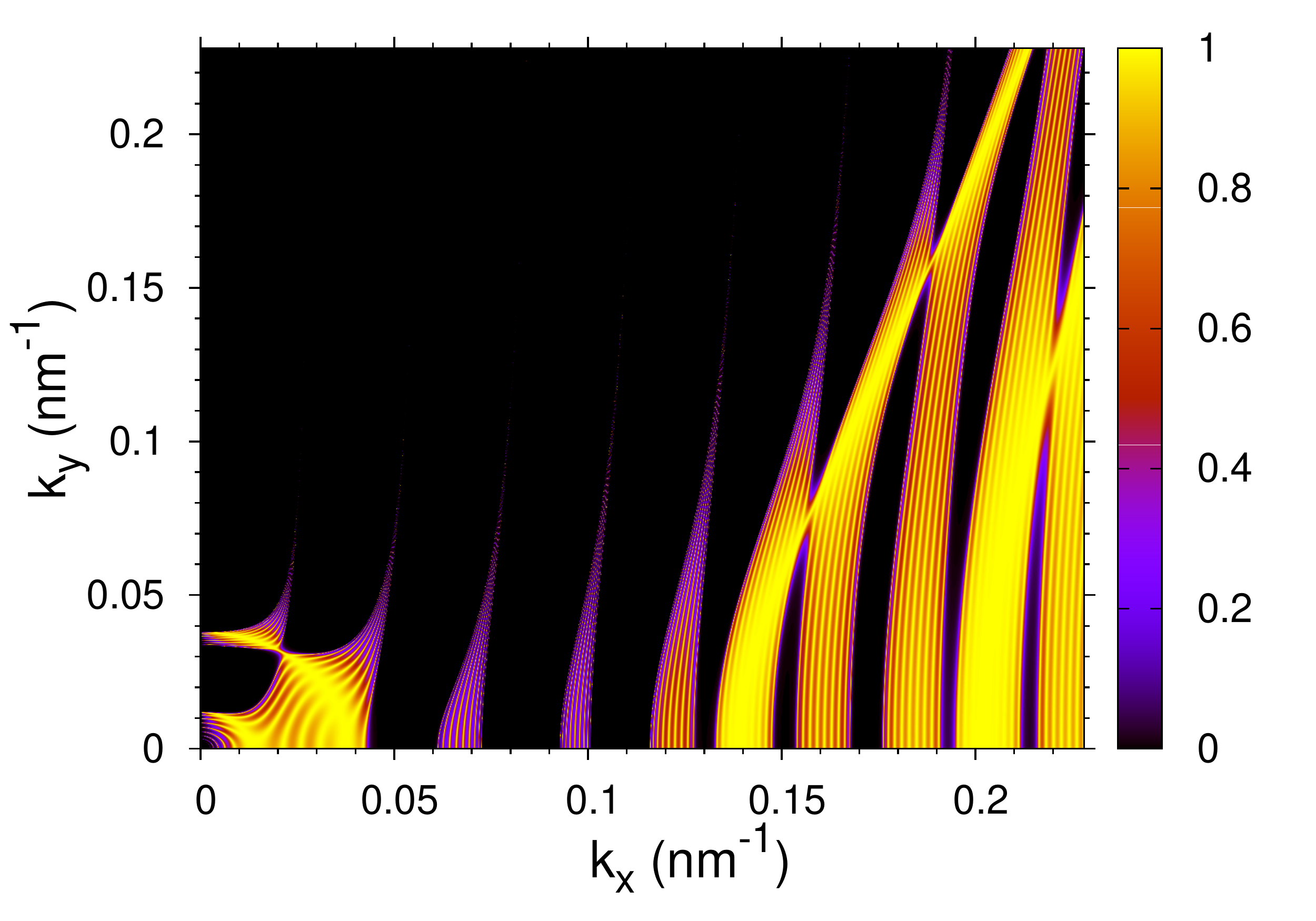}
	\end{center}
	\vspace*{-0.4cm}
	\caption{(Color online) Contour plot of the transmission of relativistic bosons in a structure consisting of ten units with $c = 10^6$ m/s, $V = 50$ meV, $a =50$ nm, $ b = 100$ nm, and $m = 0$.} \label{fig10}
\end{figure}

\begin{figure}[ht]
	\begin{center}
		\includegraphics[width=9cm]{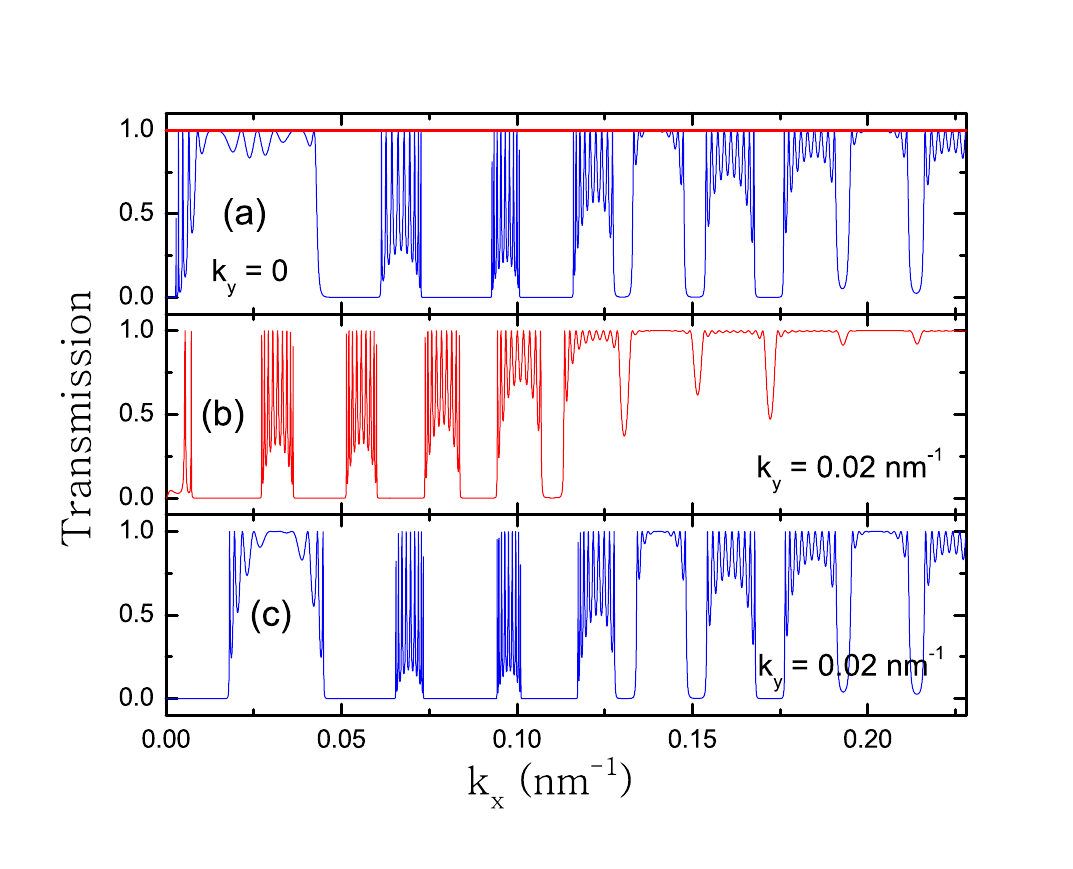}
	\end{center}
	\vspace*{-0.4cm}
	\caption{(Color online) Electrons vs bosons. Plots of the transmission through a structure consisting of ten units with $c = 10^6$ m/s, $V = 50$ meV, $a =50$ nm, $ b = 100$ nm, and $m = 0$.} \label{fig11}
\end{figure}

Again we make contact with Sec.\ II by considering the transmission through a ten-unit structure described above for fermions. A contour plot of the transmission is shown in Fig.\ \ref{fig10}. Contrasting Figs.\ \ref{fig2} and \ref{fig10} we see that the main effect of having many units is to reduce the continuous range of $k_x$ for which a perfect transmission occurs, by "inserting" small gaps between  regions of perfect transmission, and introduce resonances in the gap of Fig.\ \ref{fig2} which occurs for $0.5\leq k_x\leq 0.014$. This was not the case for fermions for which the transmission was perfect in this range.
Another way to contrast the results of the two superlattice cases considered above is shown in Fig.\ \ref{fig11} where  the transmission is plotted vs $k_x$ for ten units and  constant $k_y$. Panel (a) is for normal incidence, that is for $k_y = 0$, the red (straight) curve being for electrons and the blue one for bosons. Panels (b) and (c) are for electrons and  bosons, respectively, and in either case for $k_y = 0.02$ nm$^{-1}$. We see again the perfect transmission for electrons in panel (a) and the quite different one for bosons while for an oblique incidence the difference is less pronounced. Notice the quite "square-wave" character of the transmission, which becomes more pronounced upon increasing the number of units.  Notice also the similarity between Figs.\ \ref{fig4} and \ref{fig11}. Again the differences between electrons and  bosons are due to the chirality of the former.

\subsection{Density of states}

An additional way to contrast electrons with bosons is to evaluate the density of states (DOS) $D(E)$. In the reduced-zone scheme it is given by
\begin{equation}
	D(E)  = \frac{4A}{\pi^2} \sum_n \int_{0}^{\pi/l} d k_x \int_{0}^{\infty} d k_y \delta(E - E_n(k_x,k_y)),
\end{equation}
where $A$ is the surface area. The  integral is evaluated numerically by converting it to a sum in the manner
\begin{equation}
	\int_{0}^{\pi/l} \ud k_x \int_{0}^{\infty} \ud k_y \approx \big(\frac{\pi}{N_x l}\big) \big( \frac{C}{N_y}  \big) \sum_{k_x = 0}^{\pi/l} \sum_{k_y = 0}^{C},
\end{equation}
where  the $k_x$ and $k_y$ indices take the values
\begin{equation}
	k_x = \frac{n_x}{N_x}\frac{\pi}{l}, \,\, k_y = \frac{n_y}{N_y}C, \,\, n_x(n_y) = 1 \cdots N_x(N_y).
\end{equation}
The cutoff C in the $k_y$ direction is chosen sufficiently large so that there is negligible change in the DOS by not considering contributions from  $k_y>C$;  for the plots shown we took $C=10 \pi/l$. In addition, we  replace the $\delta$ function in Eq.\ (33) by a gaussian, of width $\sigma = 3 meV$,
\begin{equation}
\begin{aligned}
	\delta(E-E_n(k_x, k_y)) & \approx \frac{1}{\sigma\sqrt{2\pi}}
	\ e^{ -	[E-E_n(k_x, k_y)]^2/2\sigma^2},\\
\end{aligned}
\end{equation}
and  chose $\sigma$ small but  sufficiently large to compensate for the discretization of  $k_x$ and $k_y$.  For the DOS of zero-spin bosons we divide the result  by 4 since we don't have the fourfold valley-and-spin degeneracy. We also evaluate $D(E)$ for a non-relativistic superlattice in the tight-binding model
\begin{equation}
	E_n(k_x,k_y) = E_n +\hbar^2 k_y^2/2m - 2 t_n \cos(k_x l),
\end{equation}
where $E_n$ is the middle  of the $n$th miniband and $t_n$ the hopping parameter. Notice that we include the free motion along the $y$  direction. Then $D(E)$ takes the form
\begin{equation}
	D(E) = \frac{\sqrt{2m}A}{\pi h} \sum_n
	  \int_{-\pi/l}^{\pi/l} \frac{\ud k_x}{\sqrt{E - E_n + 2 t_n \cos(k_x l)}}
\end{equation}
and evaluate the integral numerically  with $t_n$ equal to one quarter of the width of the $n$-th miniband (the parameters $E_n$ and $t_n$ we obtained numerically for the same shape of superlattice as for the relativistic particles).

\begin{figure}[ht]
	\begin{center}
	\includegraphics[width=9cm]{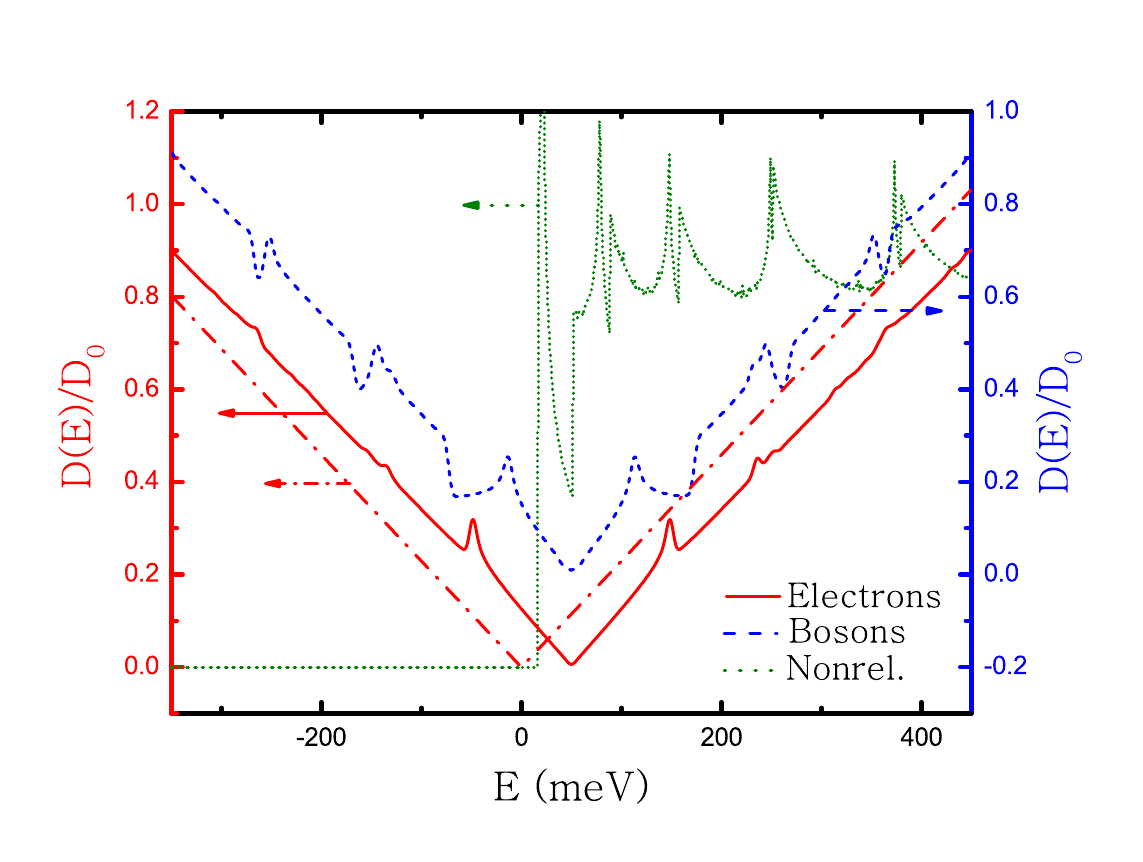}
	\end{center}
	\vspace*{-0.45cm}
	\caption{(Color online) Electrons vs bosons. Density of states $D(E)$ vs energy $E$ for a superlatice with $c = 10^6$ m/s, $V = 100$ meV, $a = b = 10$ nm, $m = 0$. The red, solid curve is for electrons and the blue, dashed one for bosons, the latter is shifted for clarity. The green dotted curve is for a non-relativistic particle ($m = 0.067m_0$)  in the tight-binding model, see Eq.\  (38), and the red dot-dashed curve for the massless Dirac electron in the absence of a superlattice. } \label{fig12}
\end{figure}

The results are shown in Fig.\ \ref{fig12}. The solid red curve is for electrons ($D_0=4A/\pi^2$) and the dashed blue one for bosons; the left and right scales are different due to the absence of spin degeneracy in the latter case. The dotted green curve is for non-relativistic electrons in the superlattice in the tight-binding model with $D_0= \sqrt{2m}A/\pi h$ and $m = 0.067m_0$. It appears different from the usual 1D one because we included the motion along the $y$ direction but not the one along the $z$ direction\cite{bas}. For comparison we also show the DOS for the Dirac 2DEG in the absence of a superlattice by the red dot-dashed curve. The structure seen in the DOS is directly related to that of the dispersion relation shown in Figs.\ \ref{fig5}, \ref{fig6}, \ref{fig8}, \ref{fig9}. The more pronounced structure in the case of bosons reflects the wider gaps, on the average, at the zone boundaries. In general the DOS increases upon approaching the band edges and then decreases. Accordingly, there are no peaks in the DOS for a Dirac 2DEG, no states with $E < 0$ for  a non-relativistic particle, etc.

\section{summary}

We obtained  the dispersion relation for massless fermions  and  zero-spin bosons using, respectively, the Dirac and Klein-Gordon equation with a 1D periodic potential but allowing for motion of these particles in two dimensions.  In doing so we generalized the treatment of Ref.\ \onlinecite{mac} which considered motion only along one direction. We also  evaluated the transmission through such a superlattice and  related  it to that through a  single barrier. Further, we evaluated the density of states for either superlattice, for a non-relativistic particle, and for the Dirac massless fermion.

The parameters we chose for  fermions are appropriate to graphene that is currently studied intensively. It is well established that electrons in graphene behave effectively as chiral, "relativistic" massless ($m = 0$) particles  with a "light speed" equal to their Fermi velocity. Though the Dirac and Klein-Gordon equations differ significantly, we chose the same parameters for zero-spin bosons for a more meaningful contrast between the results derived from them. The expression for the wavevector $K$, Eq.\ (2), indicates that taking $m\neq 0$ is equivalent to assuming a higher effective $k_y$, through  ${k^*_y}^2 = m^2 c^2 / \hbar^2 + k_y^2$ or a higher effective mass through  ${m^*}^2 = \hbar^2 k_y^2 / c^2 + m^2$. Incidentally, as we pointed out after Eq.\ (33), taking $m\neq 0$ gives a finite group velocity $v_{k_y}$ in the Klein-Gordon case, see Figs.\ \ref{fig8} and \ref{fig9}. We notice in passing, that despite infinite velocities encountered in problems studied with the Klein-Gordon equation, the equation is quite versatile as it can also describe waves in plasmas, torsion-coupled pendula, pulse propagation in periodic dielectric structures as well as tunneling in waveguides of photonic band-gap materials, see Ref.\ \onlinecite{ste} and references cited therein.

As expected, we saw that Klein tunneling \cite{klein,kat,milton1},  the perfect transmission of carriers, upon normal incidence ($k_y\neq 0$),  occurs not only through a  single potential barrier but also through a superlattice, see Figs.\ \ref{fig3} and \ref{fig11}.  As pointed out this difference from the non-relativistic case occurs because the wavevector $K$ does depend on $k_y$ in the former case but not in the second. For zero-spin bosons this perfect transmission occurs for a highly restricted range of wavevectors, cf.\ Figs.\ \ref{fig2}, \ref{fig10}, and \ref{fig11}. The difference though in the transmission between fermions  and  zero-spin bosons  is progressively less pronounced if the incidence is oblique ($k_y\neq 0$), see Figs.\ \ref{fig2}-\ref{fig4} and \ref{fig11}. Apart from these differences though, it is essential to emphasize and realize that   a linear spectrum is not an imperative requirement in order to have perfect transmission through  potential barriers, for  normal incidence, i.e., $T=1$, whereas chirality is indeed essential, cf.\ Figs.\ \ref{fig2}-\ref{fig4} and \ref{fig11}.

Finally, we saw the various DOS contrasted in Fig.\ \ref{fig12}. Notice that in the DOS  for a non-relativistic superlattice we retained the motion along the y direction, cf.\ Eq.\ (38). Apart from being meaningful or instructive, the differences between the various cases shown  will be reflected in the transport coefficients, postponed to a future study, and hopefully of use to pertinent experiments.

\section*{ACKNOWLEDGEMENTS}
 This work was supported by the Flemish Science Foundation (FWO-Vl), the Belgian Science Policy (IAP), the Brazilian council for research (CNPq) and the Canadian NSERC Grant No. OGP0121756.
\

\section{Appendix}

Below we evaluate the transmission for electrons. The solution to Eq.\ (5) is a linear combination of two independent solutions $\phi_a(x)$ and $\phi_b(x)$.  Outside the barrier we have
\begin{equation} \label{vlakkegolf1}
    {\bf \Omega_{k_x}(x)}
    = \matt{1}{1}{\lambda_+}{-\lambda_-}N_{k_x} e^{ik_xx\sigma_z},
\end{equation}
where $N_{k_x}$ is a normalization factor, $\lambda_\pm$ and $\Lambda_\pm$ are defined as
\begin{equation}
    \lambda_\pm = \frac{k_\pm\hbar c}{E' + m c^2}\text{,}\quad
    \Lambda_\pm = \frac{K_\pm\hbar c}{|E'| + m c^2},
\end{equation}
and ${\bf \Omega}_{k_x}(x)$ is given by Eq.\ (15). Inside the barrier
\begin{equation}
\left\{ \begin{aligned}
    &{\bf \Omega}_K(x) = \matt{1}{1}{\Lambda_+}{-\Lambda_-} N_Ke^{iKx\sigma_z}, \quad E>V\\
    &{\bf \Omega}_K(x) = \matt{-\Lambda_-}{\Lambda_+}{1}{1} N_Ke^{iKx\sigma_z} \quad E<V.
\end{aligned} \right.
\end{equation}
We now use the transfer matrix method and  express the solutions in terms of $\phi_a$ and $\phi_b$. Referring to Fig.\ \ref{fig1} we have
\begin{equation}
\left\{ \begin{aligned}
    \psi(x) & = {\bf \Omega}_{k_x}(x) \kvecje{A}{B} \,\,\,\,\,\,\qquad \text{in region I,}\\
    \psi(x) & = {\bf \Omega}_K(x) \kvecje{C}{D} \,\,\,\,\, \, \qquad \text{in region II,}\\
    \psi(x) & = {\bf \Omega}_{k_x}(x) \kvecje{F}{G=0} \qquad \text{in region III}.\\
\end{aligned} \right.
\end{equation}
The wave function has to be continuous at $x = 0$ and $x = W$. This gives \vspace*{-0.2cm}
\begin{equation}
    \kvec{A}{B} = {\bf \Omega}_{k_x}^{-1}(0) {\bf \Omega}_{K}(0) \kvec{C}{D},
\end{equation}
\vspace{0.01cm}
\begin{equation}
	\kvec{C}{D} = {\bf \Omega}_{K}^{-1}(W) {\bf \Omega}_{k_x}(W) \kvec{F}{G=0},
\end{equation}
and \vspace*{-0.5cm}
\begin{equation}
    \kvec{A}{B} = {\bf T} \kvec{F}{0},
\end{equation}
where ${\bf T}$ is defined as
\begin{equation}
    {\bf T} = {\bf \Omega}_{k_x}^{-1}(0) {\bf \Omega}_{K}(0) {\bf\Omega}_{K}^{-1}(W) {\bf \Omega}_{k_x}(W).
\end{equation}
Explicitly we obtain the following results for the matrix elements $T_{mn}$ of the
$2\times 2$ transfer matrix   ${\bf T}$.
For $E > V$ or  $E < 0$ the results are
\begin{equation}
\begin{aligned}
	T_{11} & =
	\alpha e^{ik_xW}\left[ |\lambda^* + \Lambda|^2 e^{-iKW} - |\lambda - \Lambda|^2 e^{iKW} \right],\\
	T_{12}&= 2i\alpha e^{-ik_xW}(\lambda- \Lambda) (\lambda+ \Lambda^*)\sin(KW),\\
	T_{21}  & = -T^*_{12},\quad 	T_{22}   = T^*_{11} ,  	
\end{aligned}
\end{equation}

\vspace*{0.05cm}
\hspace*{-0.3cm}where $\alpha =dD/4k_xK$. For  $0 < E < V$ the results are
\begin{equation}
\begin{aligned}
	T_{11}  & =
	\alpha e^{ik_xW}\left[- |\lambda\Lambda-1|^2 e^{-iKW} + |\lambda \Lambda^*+1|^2 e^{iKW} \right],\\
	T_{12}&= 2i\alpha e^{ik_xW}(-\lambda^*\Lambda^*+1) (\lambda^* \Lambda+1)\sin(KW),\\
	T_{21}  & = -T^*_{12},\quad 	T_{22}   = T^*_{11}.
\end{aligned}
\end{equation}

With reference to Fig.\ \ref{fig1} the current is given by
\begin{equation}
\begin{aligned}
    j_x & = 2ecN_{k_x}^2\lambda(|A|^2 - |B|^2), \,\, \text{in region I,}\\
    j_x & = 2ecN_{k_x}^2\lambda|F|^2,  \,\qquad\qquad \text{in region III.}
\end{aligned}
\end{equation}
Then the transmission  $\mathcal{T}\equiv \mathcal{T}(k_x,k_y) $ is
\begin{equation}
    \mathcal{T} = \frac{|F|^2}{|A|^2} = \frac{1}{|T_{11}|^2}.
\end{equation}
Explicitly, for $E > V$ the transmission is given by Eq.\ (11) whereas for $0 < E < V$ it takes the form
\begin{equation} \label{trans5}
    \mathcal{T} = \left[ 4\alpha^2F'^2 \sin^2(KW)+\cos^2(KW)\right]^{-1},
\end{equation}
with $F'= (K^2+k_{y}^2)(k_x^2+k_{y}^2)/d^2D^2 + 1 + 2k_y^2/dD$ , $d = (E + m c^2)/\hbar c$, and $D = (|E-V| + m c^2)/\hbar c$. In a similar way we obtain the transmission for  $E < 0$, which is the same as that for $ E > 0$, and the  other cases listed in Secs. II and III.


\begin{references}

\bibitem{mail}E-mail: *michael.barbier@gmail.com \\
\hspace*{2cm}  + francois.peeters@ua.ac.be \\
\hspace*{2cm}  **takis@alcor.concordia.ca\\
\hspace*{2cm}
$\dagger$ joaomilton.pereira@ua.ac.be\\

\bibitem{novo3}
K. S. Novoselov, A. K. Geim, S. V. Morozov, D. Jiang, Y. Zhang, S.
V. Dubonos, I. V. Grigorieva, and A. A. Firsov, Science, {\bf 306},
666 (2004).

\bibitem{zhang}
Y. Zhang, Y. W. Tan, H. L. Stormer, and P. Kim, Nature (London) {\bf
438}, 201 (2005).

\bibitem{zheng}
Y. Zheng and T. Ando, Phys. Rev. B {\bf 65}, 245420 (2002).

\bibitem{shara}
V. P. Gusynin and S. G. Sharapov, Phys. Rev. Lett. {\bf 95}, 146801
(2005).

\bibitem{klein}
O. Klein, Z. Phys. {\bf 53}, 157 (1929).

\bibitem{kat} M. I. Katsnelson, K. S. Novoselov, A. K. Geim,
Nature Phys. {\bf 2}, 620 (2006).

\bibitem{milton1}
J. M. Pereira Jr., V. Mlinar, F. M. Peeters and P. Vasilopoulos,
Phys. Rev. B {\bf 74}, 045424 (2006).

\bibitem{been}
C. W. J. Beenakker, Rev. Mod. Phys. (to be published)

\bibitem{milton2} J. M. Pereira Jr., P. Vasilopoulos, and F.
M. Peeters, Appl. Phys. Lett. {\bf 90}, 132122 (2007).

\bibitem{zit}
J. Schliemann, D. Loss, and R. M. Westervelt,  Phys. Rev. Lett. {\bf 94} 206801 (2005);
W. Zawadzki,  Phys. Rev. B {\bf 72} 085217 (2005);
R. Winkler, U. Zulicke, and J.  Bolte, Phys. Rev. B {\bf 75} 205314 (2007)

\bibitem{crepa} J. Schwinger, Phys. Rev. {\bf 82}, 664 (1951); D. Allor, T.  D. Cohen, and D. A. McGady, 
 arXiv:cond-mat 0708.1471 (unpublished)

\bibitem{mac} B. H. J. McKellar and G. J. Stephenson, Jr., Phys. Rev. C {\bf 35},
2262 (1987).

\bibitem{grei} W. Greiner, {\it Relativistic Quantum Mechanics}, (Springer, Berlin, 1998).

\bibitem{ste} A. M. Steinberg, P. G. Kwiat, and R. Y. Chiao,  Phys.
Rev. Lett. {\bf 71}, 708  (1993);  H. G. Winful, {\it ibid}, {\bf 90}, 023901 (2003);
Y. Japha and G. Kurizki,  Phys. Rev. A {\bf 53}, 586  (1996).

\bibitem{bas} G. Bastard, {\it Wave Mechanics Applied to Semiconductor Heterostructure}, (Les Editions de Physique, Les Ulis, 1988).

\end{references}
\end{document}